\documentclass[onecolumn]{autart}    %
\usepackage{url}
\usepackage{cite}
\usepackage{amsmath,amssymb,amsfonts} %
\usepackage{graphicx}
\usepackage[dvipsnames]{xcolor}
\usepackage{arydshln}

\usepackage{tikz}
\usetikzlibrary{arrows}
\usetikzlibrary{patterns}
\usetikzlibrary{calc}
\usepackage{multicol}

\def\BibTeX{{\rm B\kern-.05em{\sc i\kern-.025em b}\kern-.08em
    T\kern-.1667em\lower.7ex\hbox{E}\kern-.125emX}}

\makeatletter
\renewcommand*\env@matrix[1][*\c@MaxMatrixCols c]{%
  \hskip -\arraycolsep
  \let\@ifnextchar\new@ifnextchar
  \array{#1}}
\makeatother

\newcommand{\vo}[1]{#1}

\newcommand{\A}{\vo{A}}
\newcommand{\B}{\vo{B}}

\newcommand{\x}{\vo{x}}
\newcommand{\y}{\vo{y}}
\newcommand{\z}{\vo{z}}
\newcommand{\w}{\vo{w}}

\newcommand{\nuu}{N_u}

\newcommand{\nx}{N_x}
\newcommand{\ny}{N_y}
\newcommand{\nz}{N_z}
\newcommand{\nw}{N_w}

\newcommand{\Htwo}{\mathcal{H}_2}
\newcommand{\Hinf}{\mathcal{H}_{\infty}}
\newcommand{\norm}[2]{\left\lVert#1\right\rVert_{#2}}

\newcommand{\Bu}{\vo{B}_u}

\newcommand{\Bw}{\vo{B}_w}

\newcommand{\Cy}{\vo{C}_y}
\newcommand{\Cz}{\vo{C}_z}

\newcommand{\Du}{\vo{D}_u}

\newcommand{\Dw}{\vo{D}_w}

\newcommand{\Gtf}{\vo{\mathcal{G}}}

\newtheorem{theorem}{Theorem}%
\newtheorem{lemma}{Lemma} %
\newtheorem{problem}{Problem} %
\newtheorem{remark}{Remark}
\newtheorem{definition}{Definition} %
\newtheorem{proposition}{Proposition} %

\newcommand{\xdot}{\dot{\vo{x}}}

\newcommand{\set}[1]{\mathcal{#1}}

\newcommand{\I}[1]{\vo{I}_{#1}}

\newcommand{\X}{\vo{X}}
\newcommand{\J}{\vo{J}}
\newcommand{\K}{\vo{K}}
\newcommand{\W}{\vo{W}}
\newcommand{\Y}{\vo{Y}}

\newcommand{\M}{\vo{M}}
\newcommand{\N}{\vo{N}}
\newcommand{\Q}{\vo{Q}}

\renewcommand{\u}{\vo{u}}

\newcommand{\sym}[1]{\mathsf{sym}\left(#1\right)}
\newcommand{\ith}{$i^{\text{th}}$ }

\newcommand{\Real}{\mathcal R}

\renewcommand{\vec}[1]{\textbf{vec}\left(#1\right)}

\newcommand{\inner}[1]{\left\langle \vo{e}\phi_i\right\rangle}

\newcommand{\eqnlabel}[1]{\label{eqn:#1}}

\newcommand{\eqn}[1]{(\ref{eqn:#1})}

\newcommand{\fig}[1]{Fig. (\ref{fig:#1})}
\newcommand{\Fig}[1]{Fig.(\ref{fig:#1})}

\DeclareMathAlphabet{\mathbfsf}{\encodingdefault}{\sfdefault}{bx}{n}

\newcommand{\Z}{\vo{Z}}

\newcommand{\C}{\vo{C}}
\newcommand{\D}{\vo{D}}

\newcommand{\domain}[1]{\set{D}}

\newcommand{\diag}{\textbf{diag}}

\newcommand{\trace}[1]{\mathsf{tr}\left( #1 \right)}

\newcommand{\row}{\mathsf{row}}
\newcommand{\col}{\mathsf{col}}

\newcommand{\flab}[1]{\label{fig:#1}}

\newcommand{\gamv}{\gamma}
\newcommand{\gamvec}{{\vo{{\Gamma}}}}

\begin{document}

\begin{frontmatter}

\title{$\mathcal{H}_2/\mathcal{H}_\infty$ Optimal Control with Sparse Sensing and Actuation}%

\author[vedang]{Vedang M. Deshpande\thanksref{tamu}}\ead{deshpande@merl.com},    %
\author[raktim]{Raktim Bhattacharya}\ead{raktim@tamu.edu}              %

\thanks[tamu]{This work was completed during doctoral studies at Texas~A\&M~University.}
\thanks[footnoteinfo]{This work was supported by the NSF grant 1762825.}
\address[vedang]{Mitsubishi Electric Research Laboratories, Cambridge, MA, USA.}  %
\address[raktim]{Department of Aerospace Engineering, Texas A\&M University, College Station, TX, USA.}             %


\begin{abstract}                          %
In this paper, we present novel convex optimization formulations for designing full-state and output-feedback controllers with sparse actuation that achieve user-specified $\mathcal{H}_2$ and $\mathcal{H}_\infty$ performance criteria. For output-feedback control, we extend these formulations to simultaneously design control laws with sparse actuation and sensing. The sparsity is induced through the minimization of a weighted $\ell_1$ norm, promoting the efficient use of sensors and actuators while maintaining desired closed-loop performance. The proposed methods are applied to a nonlinear structural dynamics problem, demonstrating the advantages of simultaneous optimization of the control law, sensing, and actuation architecture in realizing an efficient closed-loop system.
\end{abstract}
\end{frontmatter}

\section{Introduction}
In controlling high-dimensional systems, such as those encountered in fluid dynamics \cite{CHEN_ROWLEY_2011}, structural vibration \cite{hiramoto_optimal_2000}, and thermal processes \cite{deshpande_battery_LCSSCDC2021}, sparse sensing and actuation are crucial due to their inherent complexity. For example, placing a minimal set of sensors and actuators in fluid dynamics can effectively influence critical flow modes, thereby preventing the need for extensive instrumentation. Similarly, in structural vibration control, the strategic placement of actuators and sensors can target essential vibration modes, minimizing the number of devices required. In thermal processes, sparse configurations allow for efficient temperature regulation by concentrating control efforts on key regions, thus reducing the overall number of sensors and actuators while maintaining the desired system performance.

Determining the optimal placement of sensors and actuators in such complex systems and the corresponding control law is not straightforward. Often, these locations are selected ad hoc, which can potentially constrain the closed-loop performance. Consequently, an integrated formulation that simultaneously determines the placement of sensors and actuators and the control law is essential for achieving a resource-efficient and high-performant closed-loop system.

The problem of determining the sparse set of sensors and actuators for closed-loop control has been extensively studied. Existing formulations for actuator selection in $\Htwo/\Hinf$ optimal state-feedback control typically employ row-wise sparsity-inducing penalties on the controller gain matrix \cite{dhingra_admm_2014, zare_proximal_2020, munz_sensor_2014}, or element-wise sparsity for structured controller design \cite{jovanovic_controller_2016}. 
Approaches based on integer programming and search algorithms have also been developed for static output feedback control problems \cite{nugroho_simultaneous_2018}, focusing only on the stabilization of the closed-loop system. Some approaches formulate the problem from a controllability and observability perspective \cite{manohar_optimal_2020, summers_submodularity_2016}. 

Sparse sensing is also addressed similarly by employing column-wise sparsity-inducing penalties on the observer gain matrix \cite{munz_sensor_2014}. Our previous work on $\Htwo/\Hinf$ estimation with sparse sensing presents an alternative approach by applying $\ell_1$ optimization on sensor noise to eliminate unnecessary sensors in both discrete and continuous time systems with model uncertainty \cite{deshpande2023guaranteed, deshpande2021sparse, deshpande_sparseH2Hinf_LCSS2021}. A similar approach for Kalman filtering and its variants was presented in \cite{das2020optimal, das2021optimal}.

While most formulations consider sensor and actuator selection separately, few formulations jointly formulate sensor and actuator selection problems. Notably, \cite{Argha_cdc2016_structurallySparseOpfb} apply the row-wise or column-wise sparsity-inducing penalties to $\mathcal{H}_2$ output-feedback control problems. The work in \cite{singh_ifac_2018_h2hin_opfb} addresses $\mathcal{H}_2$ and $\mathcal{H}_\infty$ problems in a relaxed mixed boolean semi-definite programming setting. There have also been some efforts for simultaneous sensor and actuator selection in the Kalman filtering framework \cite{zhong_automatica_SaA2024}.

In this paper, we introduce several novel convex optimization formulations for identifying sparse actuation and sensing configurations and the control law that ensures a specified closed-loop performance in the $\mathcal{H}_2$ and $\mathcal{H}_\infty$ sense. These formulations utilize $\ell_1$ norm minimization as a sparsity-promoting technique, eliminating unnecessary sensors and actuators from a candidate set while maintaining constraints on the closed-loop performance.

The primary contributions of this paper, and how they differ from previous work, are as follows:
\begin{enumerate}
    \item \textit{Sparse Actuation for Full State-Feedback and Output-Feedback Control} -- We present a new formulation for sparse actuation in full-state feedback and output feedback control architecture to guarantee a given $\mathcal{H}_2$ and $\mathcal{H}_\infty$ performance. The formulation is based on $\ell_1$ optimization over $\mathcal{H}_2$ norm from disturbance to controller output while constraining $\mathcal{H}_2$ or $\mathcal{H}_\infty$ norm from disturbance to controlled variables. This is a new approach to induce actuation sparseness in closed-loop systems. Previous work \cite{dhingra_admm_2014, zare_proximal_2020, munz_sensor_2014, jovanovic_controller_2016} for state-feedback control promotes sparsity by inducing sparsity in the controller gain matrix. A key advantage of our method is its ability to incorporate magnitude and rate constraints directly into the control law, a capability that previous approaches lacked. The new formulations that enable this are presented in Theorems \ref{thm:h2hinf_stfb} and \ref{thm:h2h2_stfb}  for full-state feedback and Theorems \ref{thm:h2hinf_opfb} and \ref{thm:h2h2_opfb} for output feedback problems.\\

 \item \textit{Sparse Sensing \& Actuation for Dynamic Output-Feedback Control} -- We introduce new convex optimization formulations for addressing $\mathcal{H}_2$ and $\mathcal{H}_\infty$ output-feedback control problems with sparse sensing and actuation, as detailed in Theorems \ref{thm:h2_simul} and \ref{thm:hinf_simul}. Our method for inducing sparsity in sensing and actuation is conceptually similar to the approach in \cite{Argha_cdc2016_structurallySparseOpfb}, which minimizes the row and column sums of the controller's system matrices in the $\ell_1$ sense to promote sparsity. %
While the results in \cite{Argha_cdc2016_structurallySparseOpfb} focus on incorporating group level structure of actuators and sensors into the controller and are limited to $\mathcal{H}_2$ design, our formulation extends to both $\mathcal{H}_2$ and $\mathcal{H}_\infty$ closed-loop performance. Also, the formulation in \cite{singh_ifac_2018_h2hin_opfb} results in a mixed boolean semi-definite programming problem (MB-SDP), which is NP-hard and difficult to solve. While there are methods to convert MB-SDPs to convex problems, they do not simultaneously perform the globally optimal entity selection and feedback control, yielding suboptimal results. In contrast, our convex optimization formulation ensures a globally optimal solution.    
\end{enumerate}

{\it \textbf{Notation:}} The set of real numbers is denoted by $\Real$. Real matrices are denoted by uppercase letters, unless noted otherwise. For a real square matrix $M$, $\trace{M}$ denotes its trace, and $\sym{M}:=M+M^\top$. The $i^\text{th}$ row and $j^\text{th}$ column of an arbitrary matrix $M$ are denoted by $\row_i(M)$ and $\col_j(M)$, respectively. Asterisks ($\ast$) are used to denote the off-diagonal terms of a symmetric matrix. $\Htwo$ and $\Hinf$ norms of a transfer function $\mathcal{G}(s)$ are denoted by $\norm{\mathcal{G}(s)}{\Htwo}$ and $\norm{\mathcal{G}(s)}{\Hinf}$, respectively. 

\section{Actuator Selection}
In this section, we consider the problem of simultaneous sparse actuators and controller design. The sparseness is achieved by inducing the sparseness in the vector of $\Htwo$ norms of the transfer functions from disturbances to the individual control signals. The results for the full-state feedback are presented first, followed by the results for dynamic output feedback. 

\subsection{Static Full-State Feedback}
Let us consider the following LTI system
\begin{subequations}
  \begin{align}
    \xdot(t) &= \A\x(t) + \Bu\u(t) + \Bw\w(t), \eqnlabel{sys_proc_act} \\
    \z(t) &= \Cz\x(t) + \Du\u(t) + \Dw\w(t), \eqnlabel{sys_z_act}  \\
    \u(t) &= \K\x(t), \eqnlabel{sys_u_act} 
\end{align} \eqnlabel{ct_sys_all_act}
\end{subequations}
where, $\x\in\Real^{\nx}$ is the state-vector, $\u\in\Real^{\nuu}$ is the vector of control inputs, $\w\in\Real^{\nw}$ is the bounded signal of external disturbances that may consist of process noise and/or actuator noise. The pair $(\A,\Bu)$ is assumed to be stabilizable. 
Our aim to design a full-state feedback controller gain $\K$ in \eqn{sys_u_act} such that the output vector $\z\in\Real^{\nz}$ is bounded under the effect of disturbances $\w$, and the control input vector $\u$ is sparse. 

Note that $\u$ is the vector of all possible control inputs or actuators in the system. We can achieve a sparse actuation architecture by enforcing sparseness on $\u$.

The closed-loop system follows from \eqn{ct_sys_all_act} as
\begin{subequations}
  \begin{align}
    \xdot(t) &= (\A + \Bu\K)\x(t) + \Bw\w(t), \eqnlabel{sys_proc_act_cl} \\
    \z(t) &= (\Cz+ \Du\K)\x(t)  + \Dw\w(t). \eqnlabel{sys_z_act_cl}  
\end{align} \eqnlabel{ct_sys_all_act_cl}
\end{subequations}
Let us denote the transfer function matrix of the closed-loop system \eqn{ct_sys_all_act_cl} from $\w$ to $\z$ by $\Gtf_{\z}(s)$ which is given by
\begin{align}
    \Gtf_{\z}(s) = (\Cz+ \Du\K)\left(s\I{} -(\A + \Bu\K) \right)^{-1}\Bw + \Dw.
\end{align}
The performance criterion for the bounded $\z$ can be specified in terms of the $\Htwo$ or $\Hinf$ norm of $\Gtf_{\z}(s)$, i.e. $\norm{\Gtf_{\z}(s)}{\Htwo} < \gamma_0$ or $\norm{\Gtf_{\z}(s)}{\Hinf} < \gamma_0$ must be satisfied for a given $\gamma_0>0$.

It is assumed that the precisions of actuators are known, i.e., the matrices $\Bw$ and $\Dw$ in  \eqn{ct_sys_all_act} or \eqn{ct_sys_all_act_cl} are known. This is in contrast to the complementary observer design framework presented in our previous work \cite{deshpande_sparseH2Hinf_LCSS2021} which allows unknown sensor precisions and achieves sparse sensor selection by minimization of sensor precisions. 

In the observer design framework \cite{deshpande_sparseH2Hinf_LCSS2021}, the sensor noise does not affect the system dynamics, therefore, sparseness can be achieved by minimizing sensor precisions. On the other hand, actuator noise directly affect the system behaviour in \eqn{ct_sys_all_act_cl}. Consequently, actuator precisions may not be used as a basis for sparseness in actuation architecture. In fact, sparseness and minimizing actuator precisions (i.e., maximizing some measure of actuator noise) work against each other. As actuator noise is maximized, the controller will need more control effort for satisfying the desired performance bound and attenuating the effect of actuator noise. 
Therefore, we assume that the \textit{actuator precisions are known for designing a sparse actuation architecture}. It is noted that the \textit{optimal actuator precisions can still be determined for a given set of actuators} by introducing scaling parameters in \eqn{ct_sys_all_act_cl} similar to \cite{deshpande_sparseH2Hinf_LCSS2021}. However, the present framework does not allow co-design of sparse actuation architecture with optimal actuator precisions.

To achieve sparseness in $\u$, we will minimize the channel wise $\Htwo$ norm of the transfer function from $\w$ to $u_i$, denoted by $\Gtf_{u_i}(s)$, where $u_i$ is the \ith channel actuator channel, i.e. $\u = \begin{bmatrix} u_1 & u_2 & \cdots & u_{\nuu} \end{bmatrix}^\top$.
Let {${\row_i(\K)\in\Real^{1\times\nx}}$} be the \ith row of the matrix  $\K$.
Using \eqn{sys_proc_act} and \eqn{sys_u_act}, the transfer function $\Gtf_{u_i}(s)$ is given by
\begin{align}
    \Gtf_{u_i}(s) = \row_i(\K)\left(s\I{} -(\A + \Bu\K) \right)^{-1}\Bw.
\end{align}
Since $\Htwo$ norm characterizes the energy-to-peak ratio, smaller $\norm{\Gtf_{u_i}(s)}{\Htwo}$ implies that the \ith actuator has lesser effect or importance in controlling the system. 
Moreover, if $\norm{\Gtf_{u_i}(s)}{\Htwo}$ is sufficiently small, then the \ith actuator channel may be removed from the system.
In other words, a sparse vector of $\norm{\Gtf_{u_i}(s)}{\Htwo}$ norms is representative of a sparse actuation architecture. 
Therefore, we formulate the sparse controller design problem based on channel wise $\norm{\Gtf_{u_i}(s)}{\Htwo}$ norms, as discussed below.

Let $\sqrt{\gamv_i}\geq0$ be an upperbound on $\norm{\Gtf_{u_i}(s)}{\Htwo}$, i.e.,  
\begin{equation} \eqnlabel{gamvec_upperbound_h2norms}
\begin{aligned}
   & \begin{bmatrix} \norm{\Gtf_{u_1}(s)}{\Htwo} & \norm{\Gtf_{u_2}(s)}{\Htwo} &  \cdots & \norm{\Gtf_{u_{\nuu}}(s)}{\Htwo}\end{bmatrix}^\top  \\ 
   & \qquad \leq \begin{bmatrix} \sqrt{\gamv_1} & \sqrt{\gamv_2} & \cdots & \sqrt{\gamv_{\nuu}} \end{bmatrix}^\top =: \sqrt{\gamvec}.
\end{aligned}
\end{equation}
Thus, a sparse actuation architecture can be achieved by inducing sparseness in $\gamvec$ in \eqn{gamvec_upperbound_h2norms}.
We aim to minimize the weighted $\ell_1$ norm of the $\gamvec$, i.e.
\begin{align}
    \norm{\gamvec}{1,\vo{\rho}} := \rho^\top |\gamvec|,
\end{align}
where $\vo{\rho}>0$ is a given vector of weights and $|\cdot|$ denotes the element wise absolute value. 
Therefore, problem of controller design is formally stated as follows.

\begin{problem} \label{prob:h2hinf_stfb} The $\Hinf$ full-state feedback controller design problem is formulated as the following optimization problem
\begin{equation}
    \begin{aligned}
        &\min_{\gamvec,\K} \norm{\gamvec}{1,\vo{\rho}}, \\ \text{ subject to }
        &\norm{\Gtf_{\z}(s)}{\Hinf} < \gamma_0,  \\
        &\norm{\Gtf_{u_i}(s)}{\Htwo} < \sqrt{\gamv_i},
        \quad i=1,2,\cdots,\nuu.
    \end{aligned}
    \eqnlabel{prob_h2hinf_stfb}
\end{equation}
\end{problem}
\begin{theorem}\label{thm:h2hinf_stfb}
Solution to $\Hinf$ state-feedback design \textit{Problem \ref{prob:h2hinf_stfb}} is determined by solving the following optimization problem, and the controller gain is given by $\K = \W\X^{-1}$.
\begin{equation}\left.
\begin{aligned}
  & \min\limits_{\gamvec >0,\X>0,\W} \quad  \norm{\gamvec}{1,\vo{\rho}}    \text{ subject to } \\
& \begin{bmatrix} \sym{\A\X+\Bu\W} & \Bw & (\Cz\X+\Du\W)^\top \\
                   \ast & -\gamma_0\I{} & \Dw^\top \\
                   \ast & \ast & -\gamma_0\I{} \end{bmatrix} < 0, \\
& \sym{\A\X+\Bu\W} + \Bw\Bw^\top < 0, \\ 
&   \begin{bmatrix}
    -\gamv_i & \row_i(\W) \\
    \ast & -\X
    \end{bmatrix} < 0, \quad i=1,2,\cdots,\nuu.
\end{aligned}\right\}\eqnlabel{h2hinf_stfb_thm}
\end{equation}
\end{theorem}

\vspace{-.2in}
\begin{pf}
Using bounded real lemma \cite{yaku62, yakubovich1967method,boyd1994linear}, the condition $ \norm{\Gtf_{\z}(s)}{\Hinf}  < \gamma_0$ is equivalent to the existence of a symmetric matrix $\X>0$ such that 
\begin{align}
\begin{bmatrix} \sym{\A\X+\Bu\W} & \Bw & (\Cz\X+\Du\W)^\top \\
                   \ast & -\gamma_0\I{} & \Dw^\top \\
                   \ast & \ast & -\gamma_0\I{} \end{bmatrix} < 0, \eqnlabel{h2hinf_stfb_hinf_lmi} 
\end{align}
where $\W:=\K\X$. 
Using LMI conditions for the $\Htwo$ performance index \cite{feron1992numerical}, the condition $ \norm{\Gtf_{u_i}(s)}{\Htwo} < \sqrt{\gamv_i}$ is satisfied if the following inequalities hold.
\begin{subequations}
  \begin{align}
    \sym{\A\X+\Bu\W} + \Bw\Bw^\top &< 0 \eqnlabel{h2hinf_stfb_h2_lmi}  \\
    \trace{\row_i(\K) \ \X \ \row_i(\K)^\top} &< \gamv_i.
\end{align}
\end{subequations}
The first inequality in the previous equation is linear in terms of variable $\X$ and $\W$. Noting that $\row_i(\K) \ \X \ \row_i(\K)^\top$ 
is a scalar term, and 
\begin{align}
    \row_i(\W) = \row_i(\K\X)  = \row_i(\K)\X,
\end{align}
we rewrite the second inequality as follows. 
\begin{align*}
& \trace{\row_i(\K) \ \X \ \row_i(\K)^\top} \\ 
& \quad = \row_i(\K) \ \X\X^{-1}\X \ \row_i(\K)^\top \\
& \quad = {\row_i(\W) \ \X^{-1} \ \row_i(\W)^\top} < \gamv_i.
\end{align*}
Using Schur complement, we get the following equivalent inequality.
\begin{align}
\begin{bmatrix}
-\gamv_i & \row_i(\W) \\
\ast & -\X
\end{bmatrix} < 0. \eqnlabel{h2hinf_stfb_tr_lmi}
\end{align}
Therefore, solution to the Problem \ref{prob:h2hinf_stfb} is obtained by solving the optimization problem
$$
\min\limits_{\gamvec >0,\X>0,\W}\quad  \norm{\gamvec}{1,\vo{\rho}} \text{ subject to \eqn{h2hinf_stfb_hinf_lmi},  \eqn{h2hinf_stfb_h2_lmi}, \eqn{h2hinf_stfb_tr_lmi} }.
$$
\hfill$\qed$
\end{pf}
\vspace{-.3in}

Alternatively, the closed-loop controller performance can also be specified in terms of the $\Htwo$ norm, i.e. $\norm{\Gtf_{\z}(s)}{\Htwo} < \gamma_0$. Note that $\Dw = 0$ must be satisfied for a $\norm{\Gtf_{\z}(s)}{\Htwo}$ to be well-defined. The formal problem statement for $\Htwo$ full-state feedback controller design is stated as follows. 

\begin{problem} \label{prob:h2h2_stfb} The $\Htwo$ full-state feedback controller design problem is formulated as the following optimization problem
\begin{equation}
    \begin{aligned}
        &\min_{\gamvec,\K} \norm{\gamvec}{1,\vo{\rho}}, \\ \text{ subject to }
        &\norm{\Gtf_{\z}(s)}{\Htwo} < \gamma_0,  \\
        &\norm{\Gtf_{u_i}(s)}{\Htwo} < \sqrt{\gamv_i},
        \quad i=1,2,\cdots,\nuu.
    \end{aligned}
    \eqnlabel{prob_h2h2_stfb}
\end{equation}
\end{problem}

\begin{theorem}\label{thm:h2h2_stfb}
Solution to $\Htwo$ state-feedback design \textit{Problem \ref{prob:h2h2_stfb}} is determined by solving the following optimization problem, and the controller gain is given by $\K = \W\X^{-1}$.
\begin{equation}\left.
\begin{aligned} \eqnlabel{h2h2_stfb_thm}
  & \min\limits_{\gamvec >0,\X>0,\W, \Z>0} \quad  \norm{\gamvec}{1,\vo{\rho}}    \text{ subject to } \\
& \sym{\A\X+\Bu\W} + \Bw\Bw^\top < 0 \\ 
& \begin{bmatrix} -\Z\quad & \Cz\X+\Du\W \\
                   \ast    & -\X
\end{bmatrix} < 0, \\
& \trace{\Z} < \gamma_0^2 \\
&   \begin{bmatrix}
    -\gamv_i \quad & \W_i^\top \\
    \ast & -\X
    \end{bmatrix} < 0, \quad i=1,2,\cdots,\nuu.
\end{aligned}\right\}
\end{equation}
\end{theorem}
\begin{pf}
      The performance condition $\norm{\Gtf_{\z}(s)}{\Htwo} < \gamma_0$ is replaced with equivalent LMIs in terms of matrix variable $X>0$ using the standard results \cite{yaku62, yakubovich1967method,boyd1994linear}. The remainder of the proof is similar to that of Theorem \ref{thm:h2hinf_stfb}. \hfill$\qed$
\end{pf}

\begin{remark}
    Unlike the existing approaches \cite{dhingra_admm_2014, zare_proximal_2020, jovanovic_controller_2016, munz_sensor_2014} which induce sparseness by structural penalty on the controller gain, the hardware imposed actuator limits can be readily incorporated in the optimization problems \eqn{h2hinf_stfb_thm} and \eqn{h2h2_stfb_thm} by specifying upper bounds on $\gamma_i$, i.e., $\gamma_i \leq \gamma_{\max_i}$ or $\gamvec\leq\gamvec_{\max}$ in the vector form. 
    \hfill$\qed$
\end{remark}

\begin{remark} \label{rem:RLM}
    Re-weighted $\ell_1$ minimization technique \cite{candes_enhancing_2008} is widely used to enhance the sparseness of an $\ell_1$ minimization solution. A similar approach can be employed here to enhance sparsity in $\gamvec$ by sequentially solving the optimization problem \eqn{h2hinf_stfb_thm} or \eqn{h2h2_stfb_thm} and updating the weights vector $\rho$ after each iteration of the solution. 
    \hfill$\qed$
\end{remark}

\subsection{Dynamic Output Feedback}\label{sec:opfb_acts}
For the system given by \eqn{sys_proc_act} and \eqn{sys_z_act}, we define $\y\in\Real^{\ny}$ as the sensor measurements and assume an output feedback controller as follows.
\begin{subequations}
  \begin{align}
   \y(t) &= \Cy\x(t) + \D_{yu}\u(t) + \D_{yw}\w(t), \eqnlabel{sys_y_act} \\
    \vo{\dot{x}}_{K}(t) &= \A_{K}\x_K(t) + \B_K\y(t), \\
    \u(t)  &=\C_{K}\x_K(t) +  \D_K\y(t) , 
\end{align} \eqnlabel{opfb_controller}
\end{subequations}
where $\x_K\in\Real^{N_{K}}$ is the controller state, and controller matrices are of appropriate dimensions and to be determined.  It is assumed that $(\A,\Bu)$ is stabilizable and $(\A,\Cy)$ is detectable.
We assume that $\D_{yu}=0$ in \eqn{sys_y_act} for simplicity and without loss of generality \cite{glover_hinf_1988}.

The closed-loop system is given as follows
\begin{subequations}
  \begin{align}
   \vo{\dot{x}}_{cl}(t) &= \A_{cl}\x_{cl}(t) + \B_{cl}\w(t), \\
    \z(t)  &=\C_{cl}\x_{cl}(t) +  \D_{cl}\w(t) ,
\end{align} \eqnlabel{opfb_cl_sys}
\end{subequations}
where $\x_{cl}^\top:=[\x^\top \quad \x_K]^\top$ and the subscript \textit{cl} stands for closed-loop,  
\begin{equation} \eqnlabel{cl_matrices}
  \begin{aligned}
    \A_{cl} &  = \begin{bmatrix}
    \A + \Bu\D_{K}\Cy & \Bu\C_{K} \\
    \B_{K}\Cy & \A_{K}
    \end{bmatrix}, \\ 
    \B_{cl}   &= \begin{bmatrix}
    \Bw + \Bu\D_{K}\D_{yw} \\
    \B_{K}\D_{yw} 
    \end{bmatrix}, \\
    \C_{cl} &  = \begin{bmatrix}
    \Cz + \Du\D_{K}\Cy & \Du\C_{K}
    \end{bmatrix}, \\
    \D_{cl} &= \Dw + \Du\D_{K}\D_{yw}. 
\end{aligned}
\end{equation}
Therefore, the closed-loop transfer function from $\w$ to $\z$ is 
\begin{align}
    \Gtf_{cl} = \C_{cl}(s\I{}-\A_{cl})^{-1}\B_{cl} + \D_{cl}.
\end{align}

Let the performance criterion be specified as $\norm{\Gtf_{cl}(s)}{\Hinf} \le \gamma_0$ or $\norm{\Gtf_{cl}(s)}{\Htwo} \le \gamma_0$, for given $\gamma_0>0$. For inducing sparseness in control inputs $u$, let us again consider the transfer function from $\w$ to the \ith actuator $\u_i$, i.e.,
\begin{align}
    \Gtf_{u_i}(s) = \row_i(\tilde{\C}) (s\I{}-\A_{cl})^{-1}\B_{cl} + \D_{K}\D_{yw},
\end{align}
where 
$\tilde{\C} := \begin{bmatrix}
    \D_{K}\Cy & \C_{K}
\end{bmatrix}$.
We require that $\norm{\Gtf_{u_i}(s)}{\Htwo} < \sqrt{\gamv_i}$ and we minimize weighted $\ell_1$ norm of $\gamvec$ vector to induce sparseness similar to \textit{Problem \ref{prob:h2hinf_stfb}}. 
The problem of ${\Hinf}$ optimal output feedback controller design is stated below.

\begin{problem} \label{prob:h2hinf_opfb} The $\mathcal{H}_\infty$ output feedback controller design problem with sparse actuation is formulated as the following optimization problem
\begin{equation}
\begin{aligned}
& \min_{\gamvec, \J} \norm{\gamvec}{1,\vo{\rho}}, \\ 
\text{ subject to }
  & \norm{\Gtf_{cl}(s)}{\Hinf} < \gamma_0, \\
  &\norm{\Gtf_{u_i}(s)}{\Htwo} < \sqrt{\gamv_i}, \quad i=1,2,\cdots,\nuu.
\end{aligned}
  \eqnlabel{prob_h2hinf_opfb}
 \end{equation}
\end{problem}

The matrix of controller matrices $\J$ in \textit{Problem \ref{prob:h2hinf_opfb}} is defined as follows.
\begin{align}
    \J:=\begin{bmatrix}
    \A_{K} &  \B_{K}\\ 
     \C_{K} &  \D_{K}
    \end{bmatrix}.
\end{align}

To provide a convex formulation for solving \textit{Problem \ref{prob:h2hinf_opfb}}, we adopt the change of variables presented in \cite{Scherer_multi_objective_LMIs1997}. The optimization problem will be reformulated in terms of the following new variables
\begin{equation} \eqnlabel{opfb_change_vars}
    \begin{aligned}
    {\hat\A}_K &= \N\A_{K}\M^\top + \N\B_K\Cy\X \\ & + \Y\Bu\C_K\M^\top   + \Y(\A+\Bu\D_K\Cy)\X \\
    {\hat\B}_K &= \N\B_K + \Y\Bu\D_K \\
    {\hat\C}_K &= \C_K\M^\top + \D_K\Cy\X \\
    {\hat\D}_K &= \D_K
\end{aligned}
\end{equation}
where $\X,\Y,\M,\N$ form partitions of the Lyapunov matrix $\X_{cl}$ of the closed loop system \eqn{opfb_cl_sys}, as follows,
\begin{align} \eqnlabel{Xcl_partition}
    \X_{cl} = \begin{bmatrix}
        \Y & \N \\
        \N^\top & \star
    \end{bmatrix}, \quad 
    \X_{cl}^{-1} = \begin{bmatrix}
        \X & \M \\
        \M^\top & \star
    \end{bmatrix}, 
\end{align}
where $\star$ denotes a matrix block of appropriate dimensions. Following relationships can be readily established using the identity $\X_{cl}\X_{cl}^{-1}=\I{}$
\begin{equation} \eqnlabel{def_Pi12}
    \begin{aligned}
    \X_{cl}\Pi_1 & = \Pi_2 \\
    \text{where, } \Pi_1 := \begin{bmatrix}
        \X & \I{} \\
        \M^\top & \vo{0}
    \end{bmatrix}, 
    & \quad \Pi_2 := \begin{bmatrix}
        \I{} & \Y \\
        \vo{0} & \N^\top
    \end{bmatrix}.
\end{aligned}
\end{equation}
We reformulate the \textit{Problem \ref{prob:h2hinf_opfb}} in terms of new variables 
defined in \eqn{opfb_change_vars} and present the result as the following theorem.
\begin{theorem} \label{thm:h2hinf_opfb}
Solution to $\Hinf$ output-feedback design \textit{Problem \ref{prob:h2hinf_opfb}} is determined by solving the following optimization problem.
\begin{subequations} \eqnlabel{h2hinf_opfb_thm}
\begin{equation} 
\begin{aligned}
  & \min\limits_{\gamvec >0,\J,\X_{cl}>0} \quad  \norm{\gamvec}{1,\vo{\rho}}
  \text{ subject to } 
\end{aligned}
\end{equation}

\begin{equation} \eqnlabel{h2hinf_opfb_thm_a}
\begin{aligned}
 {\begin{bmatrix}
        \sym{\A\X + \Bu{\hat \C_K}}  & \ast & \ast & \ast \\{\hat\A_K} + (\A+\Bu{\hat \D_K}\Cy)^\top & \sym{\Y\A+{\hat \B_K}\Cy} & \ast & \ast \\
        (\Bw+\Bu{\hat\D_K}\D_{yw})^\top & (\Y\Bw+{\hat\B_{K}\D_{yw}})^\top & -\gamma_0\I{} & \ast\\
        \Cz\X+\Du{\hat\C_{K}} & \Cz + \Du{\hat\D_{K}}\Cy & \Dw+\Du{\hat\D_{K}}\D_{yw} & -\gamma_0\I{}
\end{bmatrix} < 0 }, \\
\end{aligned}
\end{equation}

 \begin{equation} \eqnlabel{h2hinf_opfb_thm_b} \left.
\begin{aligned}
&\begin{bmatrix}
        \sym{\A\X + \Bu{\hat \C_K}}  & \ast & \ast \\
        {\hat\A_K} + (\A+\Bu{\hat \D_K}\Cy)^\top & \sym{\Y\A+{\hat \B_K}\Cy} & \ast \\
        (\Bw+\Bu{\hat\D_K}\D_{yw})^\top & (\Y\Bw+{\hat\B_{K}\D_{yw}})^\top & -\I{}
\end{bmatrix} < 0, \\
&\begin{bmatrix}
        \gamv_i & \row_i({\hat\C_K}) & \row_i({\hat\D_K\Cy}) \\
         \ast & \X & \I{} \\
         \ast & \ast & \Y
\end{bmatrix} > 0,  \quad i=1,2,\cdots,\nuu \\
& {\D_K}\D_{yw} = 0.
\end{aligned}\right\}
\end{equation}
\end{subequations}
\end{theorem}
\begin{pf}
Using bounded real lemma, the condition $\norm{\Gtf_{cl}(s)}{\Hinf}  < \gamma_0$ is equivalent to the existence of a symmetric matrix $\X_{cl}$ such that 
\begin{align}
\X_{cl} > 0, \,
\begin{bmatrix} \sym{\X_{cl}\A_{cl}} & \X_{cl}\B_{cl} & \C_{cl} \\
                   \ast & -\gamma_0\I{} & \D_{cl}^\top \\
                   \ast & \ast & -\gamma_0\I{} \end{bmatrix} < 0. \eqnlabel{h2hinf_opfb_hinf_lmi} 
\end{align}
A congruence transformation of the inequality $\X_{cl} > 0$ with the matrix $\Pi_1$ gives us the following LMI,
\begin{align} \eqnlabel{XY_Lypunov_lmi}
    \begin{bmatrix}
        \X & \I{} \\
        \I{} & \Y
    \end{bmatrix} > 0.
\end{align}
After performing another congruence transformation of the second inequality in \eqn{h2hinf_opfb_hinf_lmi} with the matrix $\diag(\Pi_1, \I{}, \I{})$ followed by the change of variables using \eqn{cl_matrices}, \eqn{opfb_change_vars}, \eqn{Xcl_partition}, and \eqn{def_Pi12}, we obtain
{
\begin{align} \eqnlabel{h2hinf_opfb_pf_a}
    \begin{bmatrix}
        \sym{\A\X + \Bu{\hat \C_K}}  & \ast & \ast & \ast \\{\hat\A_K} + (\A+\Bu{\hat \D_K}\Cy)^\top & \sym{\Y\A+{\hat \B_K}\Cy} & \ast & \ast \\
        (\Bw+\Bu{\hat\D_K}\D_{yw})^\top & (\Y\Bw+{\hat\B_{K}\D_{yw}})^\top & -\gamma_0\I{} & \ast\\
        \Cz\X+\Du{\hat\C_{K}} & \Cz + \Du{\hat\D_{K}}\Cy & \Dw+\Du{\hat\D_{K}}\D_{yw} & -\gamma_0\I{}
    \end{bmatrix} < 0 .
\end{align}
}

Similarly, the condition $\norm{\Gtf_{u_i}(s)}{\Htwo} < \sqrt{\gamv_i}$ is equivalent to the existence of a symmetric matrix $\X_{cl}'>0$ such that ${\D_K}\D_{yw} = 0$ and 
\begin{equation}\eqnlabel{opfb_h2_channel_wise}
  \begin{aligned}
    \begin{bmatrix}
        \sym{\A_{cl}^\top\X_{cl}'} & \X_{cl}'\B_{cl} \\
        \ast & -\I{}
    \end{bmatrix} < 0, \,
    \begin{bmatrix}
        \gamv_i & \row_i(\tilde{\C}) \\
         \ast & \X_{cl}'
    \end{bmatrix} > 0 
\end{aligned}
\end{equation} 
Note that $\X_{cl}'$ may not be same as $\X_{cl}$. However, for controller reconstruction we require that the same Lypunov matrix satisfies all specified performance conditions, i.e., we enforce $\X_{cl}' = \X_{cl}$. 
Performing congruence transformation using $\diag(\Pi_1,\I{})$ on the first inequality in \eqn{opfb_h2_channel_wise} gives us
\begin{align}
    \begin{bmatrix}
        \sym{\A\X + \Bu{\hat \C_K}}  & \ast & \ast \\
        {\hat\A_K} + (\A+\Bu{\hat \D_K}\Cy)^\top & \sym{\Y\A+{\hat \B_K}\Cy} & \ast \\
        (\Bw+\Bu{\hat\D_K}\D_{yw})^\top & (\Y\Bw+{\hat\B_{K}\D_{yw}})^\top & -\I{}
    \end{bmatrix} < 0 .
\end{align}
Similar transformation of the second inequality in \eqn{opfb_h2_channel_wise} via $\diag(1,\Pi_1)$ gives us
\begin{align}
      \begin{bmatrix}
        \gamv_i & \row_i({\hat\C}_K) & \row_i({\hat\D_K\Cy}) \\
         \ast & \X & \I{} \\
         \ast & \ast & \Y
    \end{bmatrix} > 0, 
\end{align}
which concludes the proof. 
\hfill$\qed$
\end{pf}

The problem of ${\Htwo}$ optimal output feedback controller design is stated below.
\begin{problem} \label{prob:h2h2_opfb} The $\mathcal{H}_2$ optimal output feedback controller design problem with sparse actuation is formulated as the following optimization problem
\begin{equation}
\begin{aligned}
& \min_{\gamvec, \J} \norm{\gamvec}{1,\vo{\rho}}, \\ 
\text{ subject to }
  & \norm{\Gtf_{cl}(s)}{\Htwo} < \gamma_0, \\
  &\norm{\Gtf_{u_i}(s)}{\Htwo} < \sqrt{\gamv_i}, \quad i=1,2,\cdots,\nuu.
\end{aligned}
  \eqnlabel{prob_h2h2_opfb}
 \end{equation}
\end{problem}
\begin{theorem} \label{thm:h2h2_opfb}
Solution to $\Htwo$ output-feedback design \textit{Problem \ref{prob:h2h2_opfb}} is determined by solving the following optimization problem.
\begin{subequations} \eqnlabel{h2h2_opfb_thm}
\begin{equation}
\begin{aligned}
  & \min\limits_{\gamvec >0,\J,\X_{cl}>0,\Y>0} \quad  \norm{\gamvec}{1,\vo{\rho}}
  \text{ subject to } \\ 
\end{aligned}
\end{equation} \\
\begin{equation}\left.
\begin{aligned}
& {\begin{bmatrix}
        \sym{\A\X + \Bu{\hat \C_K}}  & \ast & \ast \\ 
        {\hat\A_K} + (\A+\Bu{\hat \D_K}\Cy)^\top & \sym{\Y\A+{\hat \B_K}\Cy} & \ast  \\
        (\Bw+\Bu{\hat\D_K}\D_{yw})^\top & (\Y\Bw+{\hat\B_{K}\D_{yw}})^\top & -\I{} 
\end{bmatrix} < 0 ,} \\
& \begin{bmatrix}
    \X & \I{} & (\Cz\X + \Du{\hat\C_K})^\top \\
    \ast & \Y & (\Cz + \Du{\hat\D_K\Cy})^\top\\
    \ast & \ast & \Q
\end{bmatrix} > 0,\\ 
& \trace{\Q}<\gamma_0^2, \quad  \Dw + \Du{\hat\D_{K}}\D_{yw}=0,
\end{aligned}\right\}\eqnlabel{h2h2_opfb_thm_a}
\end{equation} \\
\begin{equation}\left.
\begin{aligned}
&\begin{bmatrix}
        \sym{\A\X + \Bu{\hat \C_K}}  & \ast & \ast \\
        {\hat\A_K} + (\A+\Bu{\hat \D_K}\Cy)^\top & \sym{\Y\A+{\hat \B_K}\Cy} & \ast \\
        (\Bw+\Bu{\hat\D_K}\D_{yw})^\top & (\Y\Bw+{\hat\B_{K}\D_{yw}})^\top & -\I{}
\end{bmatrix} < 0, \\
&\begin{bmatrix}
        \gamv_i & \row_i({\hat\C}) & \row_i({\hat\D_K\Cy}) \\
         \ast & \X & \I{} \\
         \ast & \ast & \Y
\end{bmatrix} > 0,  \quad i=1,2,\cdots,\nuu \\
& {\D_K}\D_{yw} = 0.
\end{aligned}\right\}\eqnlabel{h2h2_opfb_thm_b}
\end{equation} 
\end{subequations}
\end{theorem}
\begin{remark}
    Constraints ${\D_K}\D_{yw} = 0$ and {$\Dw + \Du{\hat\D_{K}}\D_{yw}=0$} in \eqn{h2h2_opfb_thm} require that $\Dw=0$ in \eqn{sys_proc_act}. 
\end{remark}
\begin{pf}
The condition $\norm{\Gtf_{cl}(s)}{\Htwo}  < \gamma_0$ is equivalent to the existence of a symmetric matrix $\X_{cl}$ such that 
\begin{equation} \eqnlabel{h2h2_opfb_h2_lmi} 
    \begin{aligned} 
        & \begin{bmatrix}
            \X_{cl} & \ast \\ 
            \C_{cl} & \Q
        \end{bmatrix} > 0 , \, 
        \begin{bmatrix} 
            \sym{\X_{cl}\A_{cl}} & \X_{cl}\B_{cl} \\
            \ast & -\I{}  \\
        \end{bmatrix} < 0 \\ 
        & \X_{cl} > 0, \quad \trace{\Q} < \gamma_0^2, \quad \D_{cl} = 0. 
    \end{aligned}
\end{equation}
A congruence transformation of the first two inequalities in \eqn{h2h2_opfb_h2_lmi} with the matrix $\diag(\Pi_1, \I{})$ followed by the change of variables using \eqn{cl_matrices}, \eqn{opfb_change_vars}, \eqn{Xcl_partition}, and \eqn{def_Pi12}, we obtain the first set of conditions \eqn{h2h2_opfb_thm_a}. The equivalent conditions  \eqn{h2h2_opfb_thm_b} for $\norm{\Gtf_{u_i}(s)}{\Htwo} < \sqrt{\gamv_i}, \quad i=1,2,\cdots,\nuu$ are identical to those in Theorem \ref{thm:h2hinf_opfb}.
\hfill$\qed$
\end{pf}

{\it Controller Construction:} Once the problems \eqn{h2hinf_opfb_thm} or \eqn{h2h2_opfb_thm} are solved to obtaina a feasible solution ${\hat\A_{K}}, {\hat\B_{K}}, {\hat\C_{K}}, {\hat\D_{K}}$, the controller can be constructed using the following inverse-transformation \cite{Scherer_multi_objective_LMIs1997}. First, determine non-singular matrices $\M, \N$ such that 
\begin{align}
    \M\N^\top = \I{} - \X\Y.
\end{align}
Since \eqn{XY_Lypunov_lmi} is implicitly satisfied
in synthesis LMIs, it can be verified by $\I{} - \X\Y$ is non-singular. 
Thus, a feasible solution for $\M, \N$ can be obtained via matrix decomposition, such as QR or LU decomposition. Then the controller is constructed as follows~\cite{Scherer_multi_objective_LMIs1997}
\begin{equation} \eqnlabel{opfb_control_contruct}
\begin{aligned} %
    \D_K &= {\hat\D_K} \\
    \C_K &= \left({\hat\C_K} - {\D_K} \Cy\X\right)\M^{-\top} \\ 
    \B_K &=  \N^{-1}\left({\hat\B_{K}} - \Y\Bu\D_K\right)\\
    \A_K &= \N^{-1}\Big[{\hat\A_K} - \N\B_K\Cy\X - \Y\Bu\C_K\M^\top - \Y(\A+\Bu\D_K\Cy)\X \Big]\M^{-\top}.
\end{aligned}
\end{equation}
\section{Simultaneous Sensor and Actuator Selection}
A convex optimization approach to address the sensor and actuator selection in output-feedback control is discussed in this section. In this approach, the sparseness is achieved by penalizing norms of certain rows or columns of the controller matrices instead of minimizing $\Htwo$ norms of the transfer function from disturbance to control $u_i(t)$. A similar approach for sparse full-state feedback has been presented in the literature \cite{dhingra_admm_2014, zare_proximal_2020, jovanovic_controller_2016, munz_sensor_2014}. We introduce the following definition and proposition for clarity in subsequent discussion.  

\begin{definition}
    A real matrix $\M$ is said to be column-sparse (row-sparse) if at least one column (row) of $\M$ is a zero vector.
\end{definition}
\begin{proposition} \label{prop:col_row_sparsity}
    Let $\M,\X,\Y$ be real matrices of appropriate dimensions. If matrix $\M$ is column-sparse, then the matrix product $\X\M$ is column-sparse. Similarly, if matrix $\M$ is row-sparse, then the matrix product $\M\Y$ is row-sparse.
\end{proposition}
The proof of Proposition \ref{prop:col_row_sparsity} can be easily verified. 

Let us again consider the system in \eqn{sys_proc_act}, \eqn{sys_z_act}, and output-feedback controller \eqn{opfb_controller}. From \eqn{opfb_controller} and Proposition \ref{prop:col_row_sparsity}, the $\u(t)$ vector is sparse if the matrix $\left[\C_{K} \quad \D_{K}\right]$ is row-sparse. Similarly, a sensor measurement $y_i$ in the vector $\y$ is not required if the $\left[\B_{K}^\top \quad \D_{K}^\top\right]^\top$ is column-sparse with its \ith column being a zero vector. 
Therefore, sparse actuation and sparse sensing are achieved by inducing row-sparseness on $\left[\C_{K} \quad \D_{K}\right]$ and column-sparseness on $\left[\B_{K}^\top \quad \D_{K}^\top\right]^\top$. 

Our approach uses $\Htwo/\Hinf$ performance conditions derived in the previous section and minimizes a sparsity-inducing cost function to achieve a structurally sparse output feedback controller. However, the performance conditions are derived in terms of the transformed variables ${\hat\A_{K}}, {\hat\B_{K}}, {\hat\C_{K}}, {\hat\D_{K}}$ in the previous section under the transformation \eqn{opfb_change_vars} and \eqn{opfb_control_contruct}. Therefore, we must verify that this transformation is sparsity-preserving. 

\begin{lemma} \label{lem:sparsity_preservation}
    If matrix $\left[\hat\C_{K} \quad \hat\D_{K}\right]$ is row-sparse and matrix $\left[\hat\B_{K}^\top \quad \hat\D_{K}^\top\right]^\top$ is column-sparse, then under the controller construction transformation \eqn{opfb_control_contruct}, the matrices $\left[\C_{K} \quad \D_{K}\right]$ and $\left[\B_{K}^\top \quad \D_{K}^\top\right]^\top$  are, respectively,  row-sparse and column-sparse. 
\end{lemma}
\begin{pf}
   From the controller construction equations \eqn{opfb_control_contruct}, we get
   \begin{align*}
       \left[\C_{K} \quad \D_{K}\right] &= \left[\left({\hat\C_K} - {\hat \D_K} \Cy\X\right)\M^{-\top} \quad {\hat \D_K} \right] \\
       &= \left[\hat\C_{K} \quad \hat\D_{K}\right] \begin{bmatrix}
           \M^{-\top} & \vo{0} \\
           \Cy\X\M^{-\top} & \I{}
       \end{bmatrix}
   \end{align*}
   and 
  \begin{align*}
   \begin{bmatrix}
       \B_{K} \\ \D_{K}
   \end{bmatrix} &= \begin{bmatrix}
       \N^{-1}\left({\hat\B_{K}} - \Y\Bu{\hat\D_K}\right) \\  {\hat \D_K}
   \end{bmatrix} \\
   &=  \begin{bmatrix}
       \N^{-1} &  \N^{-1}\Y\Bu \\
       \vo{0} & \I{}
   \end{bmatrix} 
   \begin{bmatrix}
       \hat\B_{K} \\ \hat\D_{K}
   \end{bmatrix} %
\end{align*}
   The lemma statement can be readily established using the result of Proposition \ref{prop:col_row_sparsity}.
   \hfill$\qed$
\end{pf}

In the light of Lemma \ref{lem:sparsity_preservation}, we can design a simultaneously sparse actuation and sensing framework by inducing row-sparseness and column-sparseness in matrices $\left[\hat\C_{K} \quad \hat\D_{K}\right]$ and $\left[\hat\B_{K}^\top \quad \hat\D_{K}^\top\right]^\top$ respectively. 
For notational simplicity, let us partition the matrices of interest in terms of their rows and columns as follows.
 \begin{align}
 \begin{bmatrix}
       \hat\C_{K} & \hat\D_{K}  
 \end{bmatrix} =: & 
 \begin{bmatrix}
       \bar{c}_1^\top \\  \bar{c}_2^\top \\ \vdots \\ \bar{c}_{\nuu}^\top
 \end{bmatrix}
    \end{align}
  \begin{align}
 \begin{bmatrix}
       \hat\B_{K} \\ \hat\D_{K}  
 \end{bmatrix} =: & 
 \begin{bmatrix}
       \bar{b}_1 &  \bar{b}_2 & \cdots & \bar{b}_{\ny}
 \end{bmatrix}.
 \end{align}
The performance of the controller is specified in terms of the closed-loop $\Htwo$ or $\Hinf$ norm, i.e., $\norm{\Gtf_{cl}(s)}{\Hinf} < \gamma_0$ or  $\norm{\Gtf_{cl}(s)}{\Htwo} < \gamma_0$. 
Therefore, the simultaneous sparse sensing and actuation problems can be stated as follows for given weights vectors
\begin{equation}
    \begin{aligned}
    \vo{\mu} & :=  \begin{bmatrix}
       \mu_1 &  \mu_2 & \cdots & \mu_{\nuu}
 \end{bmatrix}^\top \geq 0,  \\ 
 \vo{\nu} &:=  \begin{bmatrix}
       \nu_1 &  \nu_2 & \cdots & \nu_{\ny}
 \end{bmatrix}^\top \geq 0.
\end{aligned}
\end{equation}
Note that these weights can be updated sequentially to obtain a sparse solution, see Remark \ref{rem:RLM}.

\begin{problem} \label{prob:h2_opfb_simul} The simultaneous $\Htwo$-optimal sparse sensing and actuation in output-feedback controller design is formulated as the following optimization problem 
\begin{equation} \eqnlabel{h2_opfb_simul_prob}
    \begin{aligned}
    \min_{\hat \J} \sum_{i=1}^{\nuu} \mu_i\norm{\bar{c}_i}{2} &+ \sum_{i=1}^{\ny} \nu_i\norm{\bar{b}_i}{2} \\
    \text{ subject to } & \norm{\Gtf_{cl}(s)}{\Htwo} < \gamma_0,
    \end{aligned} 
\end{equation}
\end{problem}
 
\begin{problem} \label{prob:hinf_opfb_simul} The simultaneous $\Hinf$-optimal sparse sensing and actuation in output-feedback controller design is formulated as the following optimization problem 
\begin{equation} \eqnlabel{hinf_opfb_simul_prob}
    \begin{aligned}
    \min_{\hat \J} \sum_{i=1}^{\nuu} \mu_i\norm{\bar{c}_i}{2} &+ \sum_{i=1}^{\ny} \nu_i\norm{\bar{b}_i}{2} \\
    \text{ subject to } & \norm{\Gtf_{cl}(s)}{\Hinf} < \gamma_0,
    \end{aligned} 
\end{equation}
\end{problem}

\begin{remark}
    By setting $\vo{\mu} = \vo{0}$ or $\vo{\nu} = \vo{0}$ we may achieve sparse sensing or sparse actuation, respectively. When both $\vo{\mu}$ and $\vo{\nu}$ are non-zero, we may achieve simultaneous sparse sensing and sparse actuation.
\end{remark}

\begin{theorem}\label{thm:h2_simul}
    The solution to simultaneous sparse sensing and actuation output-feedback controller design \textit{Problem \ref{prob:h2_opfb_simul}} is determined by solving the following optimization problem    
\begin{align}
    &\min_{\hat{\J}} \sum_{i=1}^{\nuu} \mu_i\norm{\bar{c}_i}{2} + \sum_{i=1}^{\ny} \nu_i\norm{\bar{b}_i}{2} \,
   \text{ subject to \eqn{h2h2_opfb_thm_a} } 
\end{align}
\end{theorem}
\begin{pf}
   The condition $\norm{\Gtf_{cl}(s)}{\Htwo} < \gamma_0$ is equivalent to \eqn{h2h2_opfb_thm_a}, see proof of Theorem \ref{thm:h2h2_opfb}. \hfill$\qed$
\end{pf}

\begin{theorem}\label{thm:hinf_simul}
    The solution to simultaneous sparse sensing and actuation output-feedback controller design \textit{Problem \ref{prob:hinf_opfb_simul}} is determined by solving the following optimization problem    
\begin{align}
    &\min_{\hat{\J}} \sum_{i=1}^{\nuu} \mu_i\norm{\bar{c}_i}{2} + \sum_{i=1}^{\ny} \nu_i\norm{\bar{b}_i}{2} \,
   \text{ subject to \eqn{h2hinf_opfb_thm_a} } 
\end{align}
\end{theorem}
\begin{pf}
   The condition $\norm{\Gtf_{cl}(s)}{\Hinf} < \gamma_0$ is equivalent to \eqn{h2hinf_opfb_thm_a}, see proof of Theorem \ref{thm:h2hinf_opfb}. \hfill$\qed$
\end{pf}

\section{Active Control of a Tensegrity Wing}
\subsection{System Model}
\begin{figure}[h!]
\centering
\begin{tikzpicture}[scale=3]
\coordinate (N1) at (0,0);
\coordinate (N2) at ({cos(55)},{sin(55)});
\coordinate (N3) at ({cos((45)},{sin(-45)});
\coordinate (N4) at ({cos(23)},{sin(23)});

\coordinate (M1) at (0,{sin(55)});
\coordinate (M2) at ({cos(55)},{sin(-55)});
\coordinate (M3) at ({cos((45)},{sin(45)});
\coordinate (M4) at ({cos(23)},{sin(-23)});

\draw [-,very thick] (N1) -- ++(N2) node(n2){} -- ++(N3) node(n3){} -- ++(N4) node(n4){};
\draw [-,very thick] (M1) -- ++(M2) node(m2){} -- ++(M3) node(m3){} -- ++(M4) node(m4){};

\draw [-,thick,red] (N1) -- (m2) -- (n3) -- (m4) -- (n4) -- (m3) -- (n2) -- (M1);
\draw [-,thick,red] (m2) -- (n2);
\draw [-,thick,red] (m3) -- (n3);

\draw[pattern=north east lines] (0,-0.15) rectangle (-0.1,{sin(55)+0.15});

\draw [-,thin] {(M1)++(0,.15)} -- (0,-0.15);

\draw[thick, fill=white] (N1) circle(1pt); %
\draw[thick, fill=white] (n2) circle(1pt); %
\draw[thick, fill=white] (n3) circle(1pt); %
\draw[thick, fill=white] (n4) circle(1pt);

\draw[thick, fill=white] (M1) circle(1pt); %
\draw[thick, fill=white] (m2) circle(1pt); %
\draw[thick, fill=white] (m3) circle(1pt); %
\draw[thick, fill=white] (m4) circle(1pt); %

\node[] at ([xshift=1.1mm]$(M1)!0.2!(m2)$) {\small$b_1$};
\node[] at ([xshift=1.1mm]$(N1)!0.2!(n2)$) {\small$b_2$};
\node[] at ([xshift=1.1mm]$(n2)!0.3!(n3)$) {\small$b_3$};
\node[] at ([xshift=1.1mm]$(m2)!0.3!(m3)$) {\small$b_4$};
\node[] at ([yshift=-0.5mm]$(m3)!0.4!(m4)$) {\small$b_5$};
\node[] at ([yshift=0.9mm]$(n3)!0.4!(n4)$) {\small$b_6$};

\node[red] at ([yshift=1mm]$(M1)!0.5!(n2)$) {\small$c_1$};
\node[red] at ([yshift=-1mm]$(N1)!0.5!(m2)$) {\small$c_2$};
\node[red] at ([xshift=1mm]$(m2)!0.5!(n2)$) {\small$c_3$};

\node[red] at ([yshift=1mm]$(m3)!0.5!(n2)$) {\small$c_4$};
\node[red] at ([yshift=-1mm]$(n3)!0.5!(m2)$) {\small$c_5$};
\node[red] at ([xshift=1mm]$(n3)!0.5!(m3)$) {\small$c_6$};

\node[red] at ([yshift=1mm]$(m3)!0.5!(n4)$) {\small$c_7$};
\node[red] at ([yshift=-1mm]$(n3)!0.5!(m4)$) {\small$c_8$};
\node[red] at ([xshift=1mm]$(n4)!0.5!(m4)$) {\small$c_9$};

\draw[->,thick] (2.5,0.5) -- (2.5,0) node[midway,xshift=1em]{$g$};

\end{tikzpicture}

 \caption{A tensegrity cantilever structure under gravity acting downwards. The bars $(b_i)$ are shown in black, and the cables $(c_j)$ are shown in red.}
 \flab{canti}
 \end{figure}
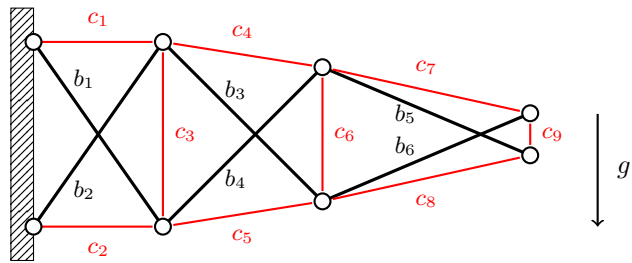
We apply the sparse actuation formulations to a tensegrity cantilever beam shown in \fig{canti}. A tensegrity structure is a network of rigid bars and elastic cables controlled by changing the force densities $\sigma$ in the cables. It is defined as $\sigma\|s\| = k(\|s\| - l_0)$ where $l_0$ is the natural length, $k$ is the spring constant, and $\|s\|$ is the cable length. In \fig{canti}, the bars are shown in black, and the cables are shown in red. The structure is clamped to the left wall, with the corresponding nodes fixed in space. The rest of the nodes are free to move. All the nodes are pin joints -- i.e., cables and bars can freely rotate around them. The structure models a morphing wing structure that is extremely light with tunable stiffness determined by the force densities in the cables. 

The degrees of freedom of the system are defined by the angles each bar makes with the horizontal. In the configuration shown, there are six bars. Therefore, the system has six degrees of freedom, resulting in a state vector $\x\in\Real^{12}$ consisting of angles and angular velocities. Also, nine cables are in the shown configuration. Therefore, the control vector is $\vo{u}\in\Real^9$. We assume disturbances act on the structure as torques on each bar; therefore, the disturbance vector is $\vo{d}\in\Real^{6}$. We assume $\|\vo{d}(t)\|_2 = 1$ and scaled by $\W_d = 0.1\I{6}$. The bars are assumed to be made of aluminum with a 5 mm radius. The cables are assumed to be 2 mm thick with Young's modulus of 0.26 GPa. The nonlinear equations of motion are derived using Lagrangian mechanics which is trimmed and linearized about the configuration shown in \fig{canti}. Further details on the modeling of tensegrity structures are available in several textbooks \cite{Oliveira2009}.

The objective is to develop a feedback control law with a minimum number of actuated cables so that the wing's shape doesn't change much under unsteady wind loading (modeled as disturbance torques on the bars). We treat this as a disturbance rejection problem with a minimum number of actuators and sensors.

We next show results from the various formulations for the $\mathcal{H}_2$ and $\mathcal{H}_\infty$ closed-loop performances. The sparse solutions were obtained using the re-weighted $\ell_1$ minimization method, see Remark \ref{rem:RLM} which is applicable to all convex formulations presented in the previous sections. 

\subsection{Sparse Actuation with Full State Feedback $\mathcal{H}_2$/$\mathcal{H}_\infty$ Control}
Here we consider the full state-feedback design to bound the perturbation in the bar angles due to the disturbances. \Fig{full_state_feedback_precisions} shows the $\|u_i(t)\|_\infty$ from the $\mathcal{H}_\infty$ and the $\mathcal{H}_2$ optimal designs for various values of $\gamma_0$, using theorems 1 and 2 respectively. \Fig{full_state_feedback_xTraj} shows the state trajectories from the  closed-loop and open-loop \textit{nonlinear} simulations. \Fig{full_state_feedback_uTraj} shows the control trajectories for the closed-loop simulations. We overlay results from the $\mathcal{H}_2$ and the $\mathcal{H}_\infty$ cases.

We observe from \Fig{full_state_feedback_precisions} that for the $\mathcal{H}_2$ and the $\mathcal{H}_\infty$ case only actuators 3, 4, 7 are necessary to achieve the desired closed-loop disturbance attenuation ($\gamma_0 = 0.1$). The associated $\|u_i(t)\|_\infty$ for the remaining actuators are very small. This is also supported by the control trajectory plots in \Fig{full_state_feedback_uTraj}. 

\Fig{full_state_feedback_precisions_gamUB} shows the $\|u_i(t)\|_\infty$ from the $\mathcal{H}_\infty$ and the $\mathcal{H}_2$ optimal designs with upper bounds on the channel-wise norms which can account for hardware imposed limits.  The optimization problems in theorems 1 and 2 were augmented with additional constraints $\gamma_i\leq81$ and $\gamma_i\leq10$ for $\mathcal{H}_\infty$ and the $\mathcal{H}_2$ design problems, respectively. We observe from \Fig{full_state_feedback_precisions_gamUB} that additional two actuators 5, 6 are needed to satisfy the performance conditions compared to the unbounded $\gamma_i$ case shown in \Fig{full_state_feedback_precisions}, thus, highlighting the trade-off between the actuation sparseness and the peak allowable control magnitude.

\begin{figure}[h!]\centering
\includegraphics[width=0.45\textwidth]{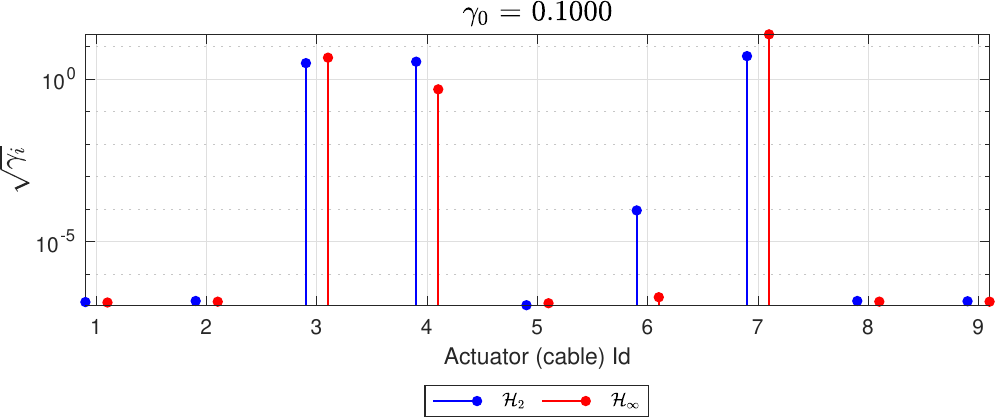}
\caption{Minimum $\|u_i(t)\|_\infty$ for a given $\mathcal{H}_2$ and $\mathcal{H}_\infty$ performance, with full-state feedback controller.}
\flab{full_state_feedback_precisions}
\end{figure}

\begin{figure}[h!]\centering
\includegraphics[width=0.45\textwidth]{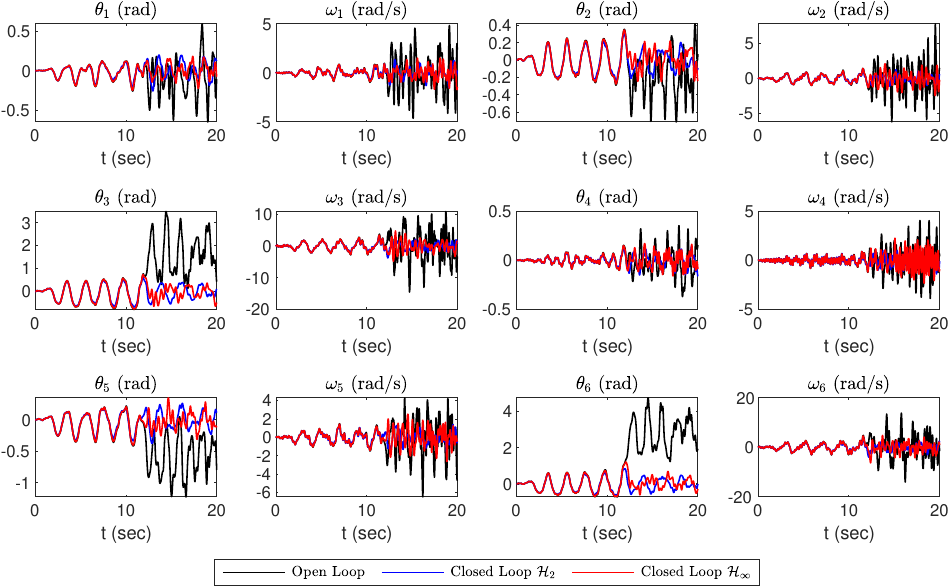}
\caption{State trajectories for a given $\mathcal{H}_2$ and $\mathcal{H}_\infty$ performance, with full-state feedback controller and nonlinear dynamics. The figure also shows the open-loop trajectories.}
\flab{full_state_feedback_xTraj}
\end{figure}

\begin{figure}[h!]\centering
\includegraphics[width=0.45\textwidth]{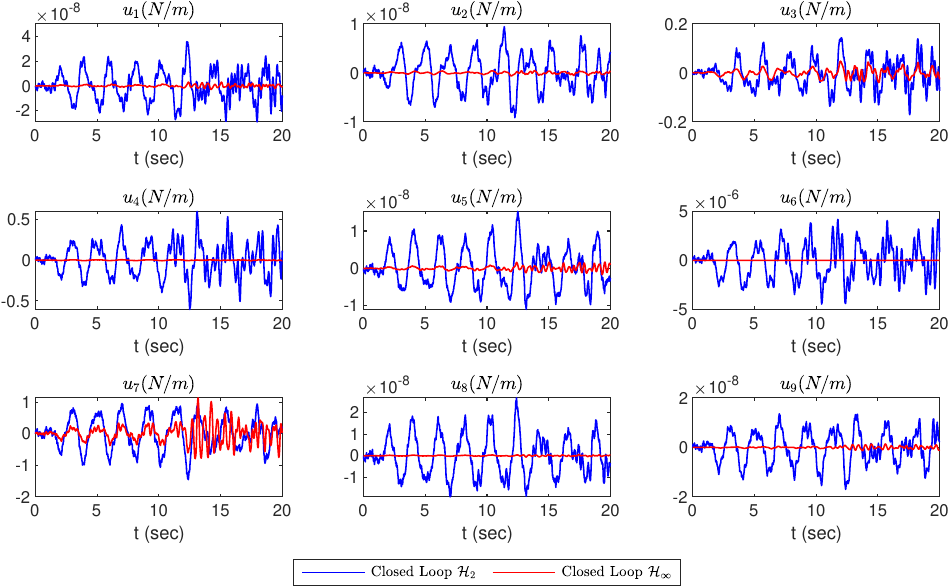}
\caption{Control trajectories for a given $\mathcal{H}_2$ and $\mathcal{H}_\infty$ performance, with full-state feedback controller and nonlinear dynamics.}
\flab{full_state_feedback_uTraj}
\end{figure}

\begin{figure}[h!]\centering
\includegraphics[width=0.43\textwidth]{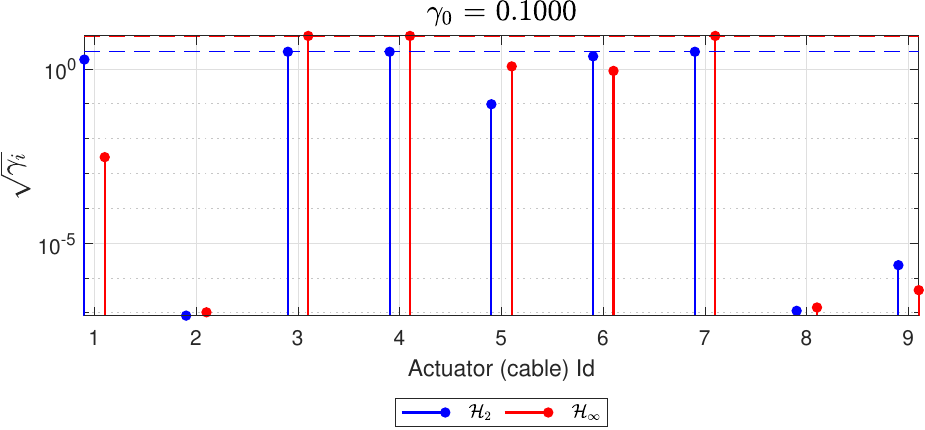}
\caption{Minimum $\|u_i(t)\|_\infty$ for a given $\mathcal{H}_2$ and $\mathcal{H}_\infty$ performance, with full-state feedback controller. The  channel-wise norms were subject to upper bounds $\gamma_i\leq10$ and $\gamma_i\leq81$ shown by dashed lines for $\mathcal{H}_2$ and $\mathcal{H}_\infty$ design problems, respectively.}
\flab{full_state_feedback_precisions_gamUB}
\end{figure}

\subsection{Sparse Actuation with $\mathcal{H}_2/\mathcal{H}_\infty$ Output Feedback Control}
Here, we consider an output-feedback $\mathcal{H}_2/\mathcal{H}_\infty$ disturbance rejection problem for the structure in \fig{canti}, with the fewest actuators. The problem is set up with measurements that include angular position and velocity for each of the 6 bars, therefore $\vo{y}\in\Real^{12} = \left[\begin{array}{cccccccccccc} \theta_1 & \omega_1 & \theta_2 & \omega_2 & \theta_3 & \omega_3 & \theta_4 & \omega_4 & \theta_5 & \omega_5 & \theta_6 & \omega_6 \end{array}\right]^\top$.

\begin{figure}[h!]\centering
\includegraphics[width=0.45\textwidth]{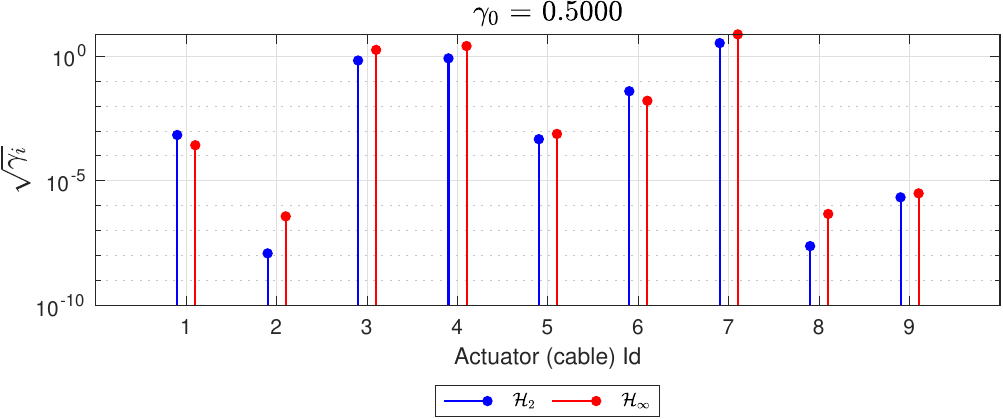}
\caption{Minimum $\|u_i(t)\|_\infty$ for a given $\mathcal{H}_2$ and $\mathcal{H}_\infty$ performance. The controller is output feedback with 12 sensors.}
\flab{precisions_channelwise_actuator}
\end{figure}

\fig{precisions_channelwise_actuator} shows the minimum $\|u_i(t)\|_\infty$  for the $\mathcal{H}_\infty$ and $\mathcal{H}_2$ formulations presented in theorems \ref{thm:h2hinf_opfb} and \ref{thm:h2h2_opfb}, respectively. In this formulation, and for this example, we observe both formulations require the same number of actuated cables with a comparable peak control magnitude. Similar to the previous case, only cables 3, 4, and 7 are actuated, with slight actuation of cable 6. 
This is supported by the control trajectories shown in \Fig{ofb_uTraj_channelwise}. A comparison of \fig{precisions_channelwise_actuator} and \fig{precisions_channelwise_actuator_gamUB} further highlight the trade-off between the sparse actuation and actuator limits. Additional actuators 1, 5, and 6 are needed to satisfy the performance conditions when the peak control magnitudes are upper-bounded as shown in \fig{precisions_channelwise_actuator_gamUB}.

\begin{figure}[h!]\centering
\includegraphics[width=0.45\textwidth]{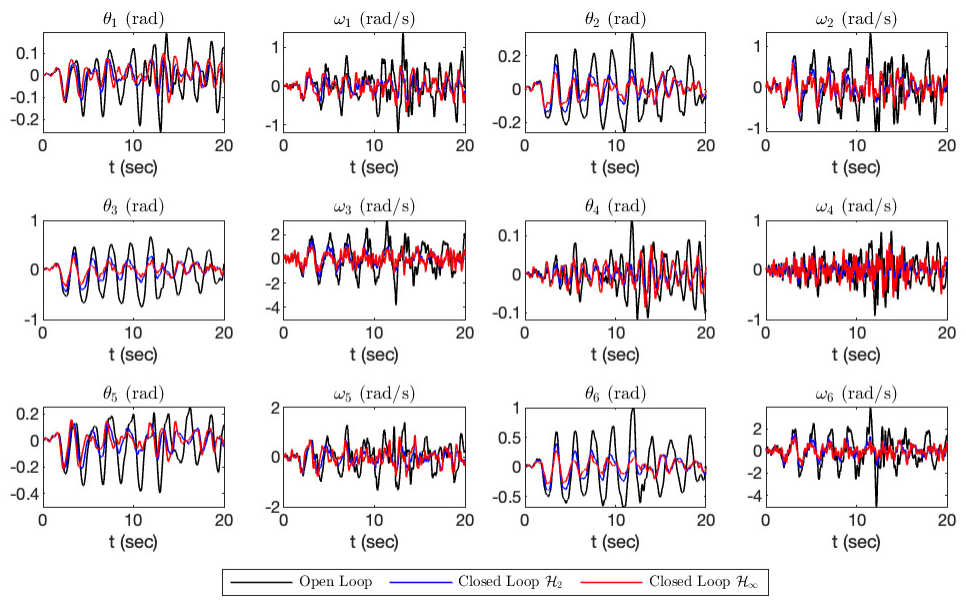}
\caption{State trajectories for a given $\mathcal{H}_2$ and $\mathcal{H}_\infty$ performance, with output feedback controller and nonlinear dynamics. The figure also shows the open-loop trajectories.}
\flab{ofb_xTraj_channelwise}
\end{figure}

\begin{figure}[h!]\centering
\includegraphics[width=0.45\textwidth]{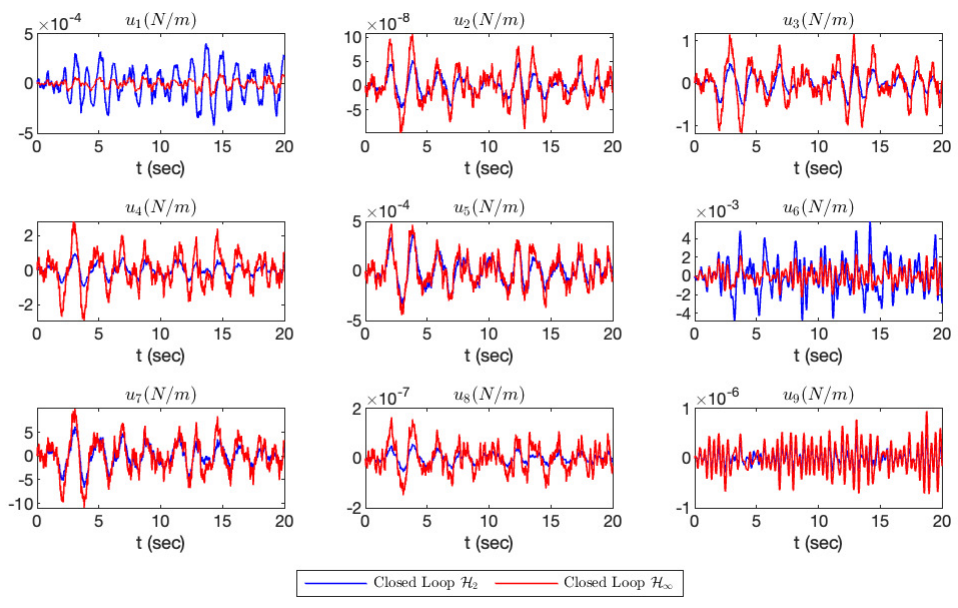}
\caption{Control trajectories for a given $\mathcal{H}_2$ and $\mathcal{H}_\infty$ performance, with output feedback controller and nonlinear dynamics.}
\flab{ofb_uTraj_channelwise}
\end{figure}

\begin{figure}[h!]\centering
\includegraphics[width=0.43\textwidth]{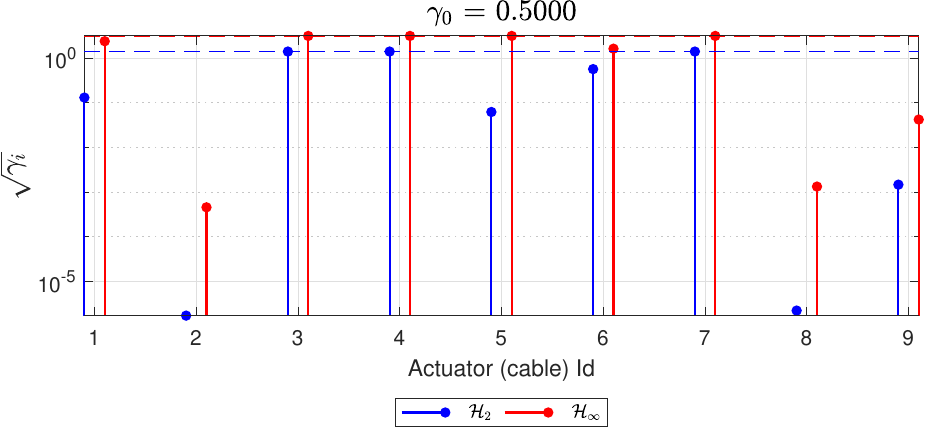}
\caption{Minimum $\|u_i(t)\|_\infty$ for a given $\mathcal{H}_2$ and $\mathcal{H}_\infty$ performance. The controller is output feedback with 12 sensors and upper bound on channel-wise norms. The channel-wise norms were subject to upper bounds $\gamma_i\leq2$ and $\gamma_i\leq10$ shown by dashed lines for $\mathcal{H}_2$ and $\mathcal{H}_\infty$ design problems, respectively.}
\flab{precisions_channelwise_actuator_gamUB}
\end{figure}

\subsection{Sparse Sensing \& Actuation with $\mathcal{H}_2/\mathcal{H}_\infty$ Output Feedback Control}
Here, we consider an output-feedback $\mathcal{H}_2/\mathcal{H}_\infty$ disturbance rejection problem for the structure in \fig{canti} with the fewest actuators and sensors. We start with the same 12 sensors and 9 actuators as in the previous case but determine the minimum number of sensors and actuators and the corresponding output feedback control law that achieves a desired closed-loop performance, using theorems \ref{thm:h2_simul} and \ref{thm:hinf_simul}.

\Fig{precisions_structurally_sparse_actuator} shows the row-norms of $\left[\hat\C_{K} \quad \hat\D_{K}\right]$ for the $\mathcal{H}_2$ and $\mathcal{H}_\infty$ formulations. Both formulations require only cables 3, 4, 6, and 7 to be actuated for $\gamma_0 = 0.1$. In this formulation, the sixth cable is significantly actuated. Also, this formulation results in larger control actions in the actuated cables, in comparison to the other two formulations. However, \Fig{precisions_structurally_sparse_sensing} indicates that only sensors 4, 6, 10, and 12 are needed for the $\mathcal{H}_2$ control and sensors 6, 10, 11 and 12 are needed for the $\mathcal{H}_\infty$ control. Therefore, there is a difference in the sensing needs for the two control objectives. The state and control trajectories of this controller are shown in \fig{xTraj_ofb_structurally_sparse} and \fig{uTraj_ofb_structurally_sparse} respectively.

\begin{figure}[h!]\centering
\includegraphics[width=0.45\textwidth]{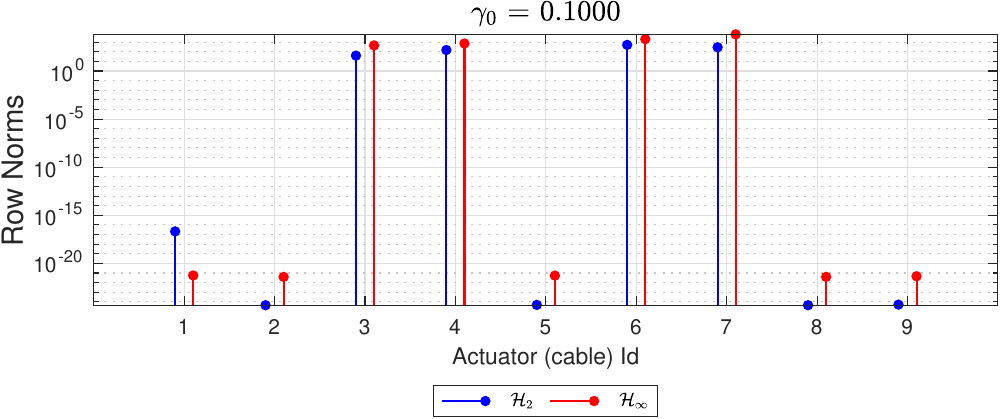}
\caption{Row norms of $\left[\hat\C_{K} \quad \hat\D_{K}\right]$ for a given $\mathcal{H}_2$ and $\mathcal{H}_\infty$ performance.}
\flab{precisions_structurally_sparse_actuator}
\end{figure}

\begin{figure}[h!]\centering
\includegraphics[width=0.45\textwidth]{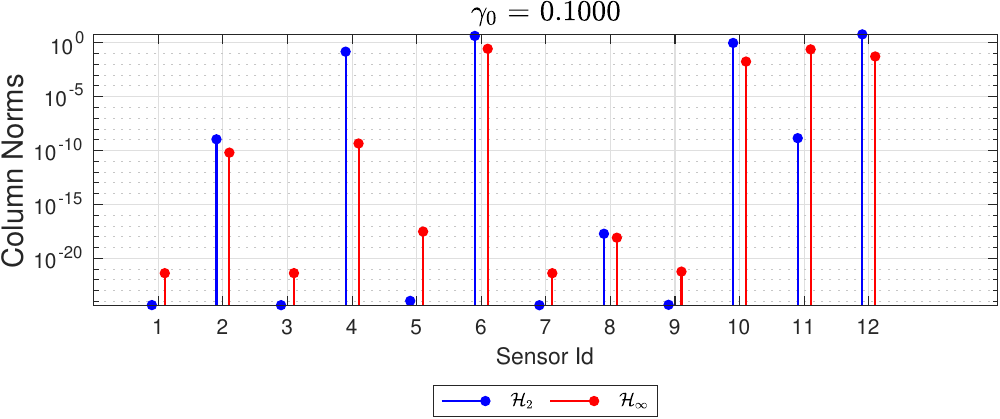}
\caption{Column norms of $\left[\hat\B_{K}^\top \quad \hat\D_{K}^\top\right]^\top$ for a given $\mathcal{H}_2$ and $\mathcal{H}_\infty$ performance.}
\flab{precisions_structurally_sparse_sensing}
\end{figure}

\begin{figure}[h!]\centering
\includegraphics[width=0.45\textwidth]{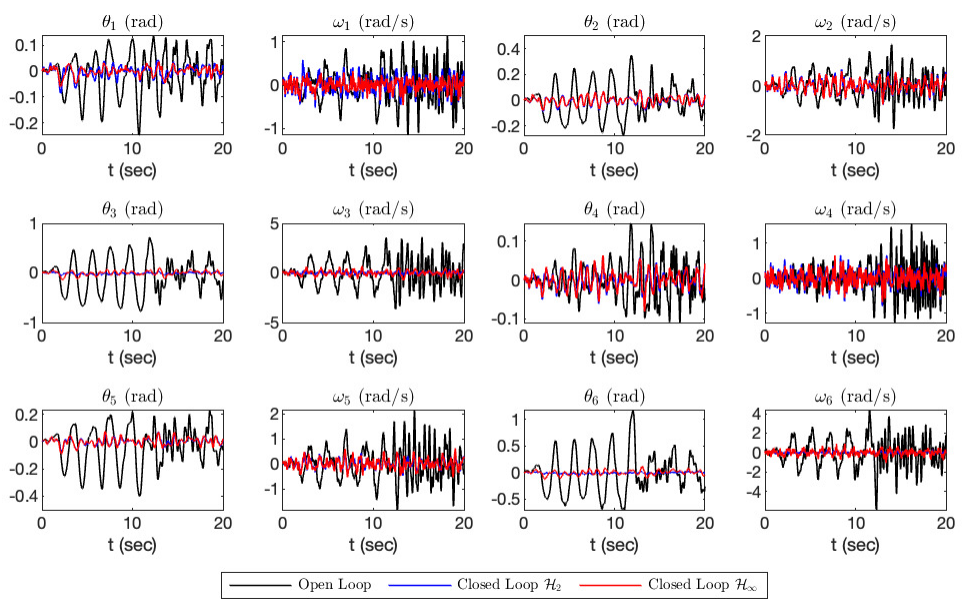}
\caption{State trajectories for a given $\mathcal{H}_2$ and $\mathcal{H}_\infty$ performance, with output feedback controller and nonlinear dynamics. The figure also shows the open-loop trajectories.}
\flab{xTraj_ofb_structurally_sparse}
\end{figure}

\begin{figure}[h!]\centering
\includegraphics[width=0.45\textwidth]{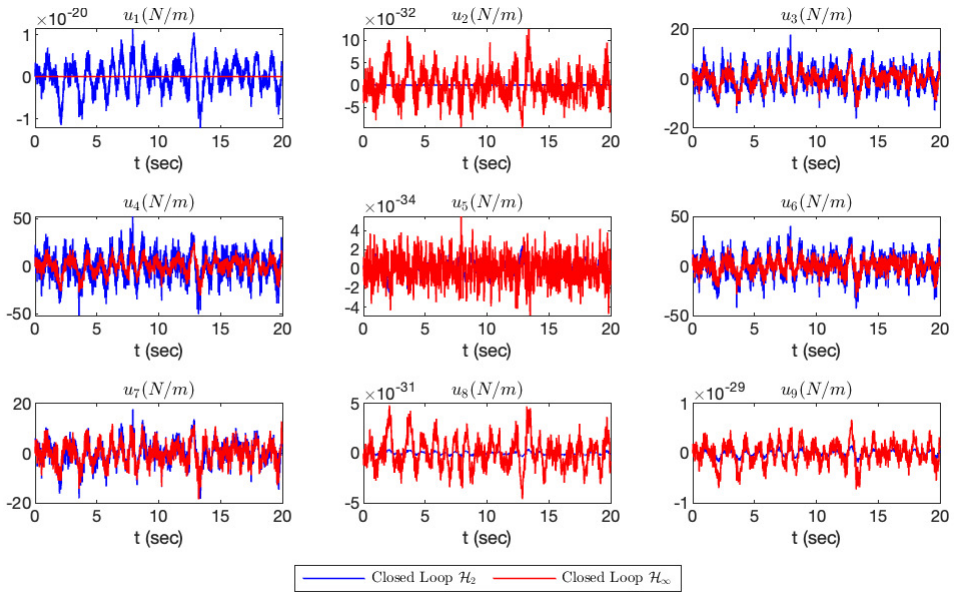}
\caption{Control trajectories for a given $\mathcal{H}_2$ and $\mathcal{H}_\infty$ performance, with output feedback controller and nonlinear dynamics.}
\flab{uTraj_ofb_structurally_sparse}
\end{figure}

\section{Summary \& Conclusions}
In conventional control system design, sensors and actuators are typically selected based on open-loop considerations, without accounting for the interaction with the control law. Once these components are determined, the control law is then designed, which may result in a system that either fails to meet the closed-loop objectives or includes unnecessary sensors and actuators. The concept of simultaneously determining the minimum number of sensors and actuators alongside the control law is therefore intriguing, as it enables more efficient system realization by ensuring that only the essential components are used while still achieving the desired closed-loop performance.

In this paper, we have developed novel convex optimization formulations for the design of full-state and output-feedback controllers with sparse actuation, tailored to meet user-specified $\mathcal{H}_2$ and $\mathcal{H}_\infty$ performance requirements. For output-feedback control, we further extended these formulations to enable the simultaneous design of control laws with sparse actuation and sensing. The sparseness is achieved by formulating the control design problem with several feasible actuators and sensors and minimizing a weighted $\ell_1$ norm of suitably defined quantities with constraints defining the closed-loop performance. The redundancy in the actuation and sensing potentially introduces a sparse solution. The $\ell_1$ optimization eliminates unnecessary components while meeting the closed-loop performance. The effectiveness of these methods was demonstrated through their application to a nonlinear structural dynamics problem, highlighting the benefits of optimizing control, sensing, and actuation architecture simultaneously to achieve an efficient system realization.
Also, the algorithms presented in this paper identify the critical sensors and actuators required to achieve the desired closed-loop performance in sparse actuation and sensing architectures. Since the failure or degradation of any critical component could compromise the controller's ability to maintain the specified performance, the results can be used to systematically determine which sensors and actuators require higher quality or redundancy to meet reliability objectives.

\bibliographystyle{unsrt}        %
\bibliography{MyLibrary}           %
\end{document}